\documentstyle[12pt]{article}
\def\be{\begin{equation}}
\def\ee{\end{equation}}
\def\l{\label}
\def\C{{\cal C}}

\def\S{{\cal S}}

\def\T{{\cal T}}
\def\V{{\cal V}}
\def\Z{{}}
\def\H{{\cal H}}
\def\O{{\cal O}}

\def\U{{\cal U}}
\def\W{{\cal W}}
\def\Q{{\cal Q}}
\def\V{{\cal V}}

\def\s{{\tt s}}
\def\t{{\tt t}}

\def\NPB#1#2#3{{\it Nucl.\ Phys.}\/ {\bf B#1} (19#2) #3}
\def\PLB#1#2#3{{\it Phys.\ Lett.}\/ {\bf B#1} (19#2) #3}
\def\PRD#1#2#3{{\it Phys.\ Rev.}\/ {\bf D#1} (19#2) #3}
\def\PRL#1#2#3{{\it Phys.\ Rev.\ Lett.}\/ {\bf #1} (19#2) #3}

\def\MODA#1#2#3{{\it Mod.\ Phys.\ Lett.}\/ {\bf A#1} (19#2) #3}

\begin{document}
\begin{titlepage}

\rightline{\tt hep-th/9711028}
\rightline{UFIFT-HEP-97-33}
\rightline{DFPD97/TH/33}

\vspace{.33cm}

\begin{center} 

{\Large \bf The Equivalence Principle of Quantum Mechanics:}

\vspace{0.33cm}

{\Large \bf Uniqueness Theorem}

\vspace{1.33cm}

{\large Alon E. Faraggi$^{1}$ $\,$and$\,$ Marco Matone$^{2}$\\}
\vspace{.33in}
{\it $^{1}$ Institute for Fundamental Theory, Department of Physics, \\
            University of Florida, Gainesville, FL 32611, USA\\
            e-mail: faraggi@phys.ufl.edu\\}
\vspace{.033in}
{\it $^{2}$ Department of Physics ``G. Galilei'' -- Istituto 
            Nazionale di Fisica Nucleare\\
            University of Padova, Via Marzolo, 8 -- 35131 Padova, Italy\\
            e-mail: matone@padova.infn.it\\}

\end{center}

\vspace{0.33cm}

\centerline{\large Abstract}

\vspace{0.33cm}

\noindent
Recently we showed that the postulated diffeomorphic equivalence of states 
implies quantum mechanics. This approach takes the canonical variables to be 
dependent by the relation $p=\partial_q{\cal S}_0$ and exploits a basic 
$GL(2,C)$--symmetry which underlies the canonical formalism. In particular,
we looked for the special transformations leading to the free system with
vanishing energy. Furthermore, we saw that while on the one hand the
equivalence principle cannot be consistently implemented in classical
mechanics, on the other it naturally led to the quantum analogue of the
Hamilton--Jacobi equation, thus implying the Schr\"odinger equation.
In this letter we show that actually the principle uniquely leads to this 
solution. Furthermore, we find the map reducing any system to the free one
with vanishing energy
and derive the transformations on $\S_0$ leaving the wave function invariant.
We also express the canonical and Schr\"odinger equations by means of the
brackets recently introduced in the framework of $N=2$ SYM. These brackets are
the analogue of the Poisson brackets with the canonical variables taken as
dependent.

\vspace{0.33cm}

\noindent
{October, 28 1997}

\end{titlepage}\newpage
\setcounter{footnote}{0} 
\renewcommand{\thefootnote}{\arabic{footnote}}

It is well--known that the classical Hamilton--Jacobi (HJ) formalism stems from
the problem of finding the canonical transformation yielding a vanishing
Hamiltonian. In \cite{1} we took the canonical variables $q$ and $p$ as
dependent through the momentum generating function, that is $p=\partial_q\S_0$,
and, according to the diffeomorphic equivalence principle \cite{1}, we looked
for coordinate transformations connecting different physical systems
including the free one with vanishing energy.

The equivalence principle was suggested by a basic $GL(2,C)$--symmetry of the
canonical equation associated to the Legendre transformation of the Hamilton's
characteristic function. This connection between the Legendre transformation and
differential equations, which was used in the framework of the Schr\"odinger
equation in \cite{2}, had been introduced in \cite{3} for deriving the inversion
formula in $N=2$ super Yang--Mills, and had been further investigated in
\cite{Carroll}. The formalism naturally fits with the brackets introduced in
\cite{BOMA}. Remarkably, we can express the canonical \cite{1} and Schr\"odinger
equations in terms of these brackets that in our approach are analogous to the
Poisson brackets with the canonical variables taken as dependent.

A basic step in the construction was the proof that the equivalence principle
cannot be consistently implemented in classical mechanics. Actually, this
principle leads to the quantum analogue of the HJ equation and in turn implies
the Schr\"odinger equation \cite{1}.
We now proceed to show that the equivalence principle uniquely leads to this
solution.

Let us start with a very explicit example of the transformations we will 
consider.
Given two functions, say $f_1(x_1)=x_1^m$, $f_2(x_2)=x_2^n$, we can 
associate the
coordinate transformation $x_1\longrightarrow x_2=x_1^{m/n}$ which is
naturally induced by the identification $f_2(x_2)=f_1(x_1)$. This is 
equivalent to say that given the function $f_1(x_1)=x_1^m$, 
the map $x_1\longrightarrow x_2=v(x_1)
=x_1^{m/n}$ induces the transformation $f_1\longrightarrow f_2$, 
defined by $f_2(x_2)=f_1(x_1)$. In other words, the 
diffeomorphism $x_1\longrightarrow x_2=v(x_1)$
induces the functional transformation $f_1\longrightarrow f_2=f_1\circ v^{-1}$.

Let us now consider the case of two physical systems with Hamilton's
characteristic
functions\footnote{In literature the Hamilton's characteristic function is also
called reduced action.} $\S_0$ and $\S_0^v$. Let us denote the coordinates of 
the two systems by $q$ and $q^v$ respectively. Setting
\be
\S_0^v(q^v)=\S_0(q),
\l{is1}\ee
induces the map
\be
q\longrightarrow q^v=v(q),
\l{6xxx}\ee
where $v=\S_0^{{v}^{\;-1}}\circ \S_0$,
with $\S_0^{{v}^{\;-1}}$ denoting the inverse of $\S_0^v$. This construction is
equivalent to say that the map (\ref{6xxx}) induces the transformation $\S_0
\longrightarrow \S_0^v=\S_0\circ v^{-1}$, that is $\S_0(q)\longrightarrow\S_0^v
(q^v)=\S_0(q(q^v))$. In other words, for a given $v$ there is the induced map
${v^{-1}}^*$ defined by
$$
{v^{-1}}^*: \S_0\mapsto {v^{-1}}^*(\S_0),
$$
that is $\S_0^v(q^v)=\S_0(v^{-1}(q^v))$ so that $\S_0^v$ is the pullback of
$\S_0$ by ${v^{-1}}^*$. We will call the diffeomorphisms (\ref{6xxx})
$v$--transformations. Let us consider the Legendre transformation \cite{1}
\be
\S_0(q)=pq-\T_0(p),
\l{y1}\ee
\be
p={\partial\S_0\over\partial q},\qquad q={\partial\T_0\over \partial p}.
\l{y2}\ee

The second derivative of (\ref{y1}) with respect to $\s=\S_0(q)$ yields the
``canonical equation''
\be
\left(\partial^2_{\tt s} +\U({\tt s})\right)q\sqrt p
=0=\left(\partial^2_{\tt s}+\U({\tt s})\right)\sqrt p,
\l{10}\ee
with $\U({\tt s})=\{q\sqrt p/\sqrt p,{\tt s}\}/2=\{q,{\tt s}\}/2$, and where
$\{h(x),x\}={{h{'''}(x)}\over {h{'}(x)}}-{3\over 2}\left({h{''}(x)\over
{h{'}(x)}}\right)^2$, denotes the Schwarzian derivative. Observe that the
choice of the coordinates $q$ and $q^v$, which of course does not imply any
loss of generality as both $q$ and $q^v$ play the role of independent coordinate
in their own system, allows us to look at the reduced action as a scalar 
function.
In particular, since $\S^v_0(q^v)=\S_0(q)$, we see that the transformations
(\ref{6xxx}) leave the Legendre transformation of $\T_0$ (\ref{y1}) unchanged.
Consequently, since $\partial_{q^v}\S_0^v(q^v)=\left(\partial_q q^v\right)^{-1}
\partial_{q}\S_0(q)$, we have
\be
p\longrightarrow p_v=\left(\partial_q q^v\right)^{-1} p.
\l{cf1}\ee
However, while the Legendre transformation of $\T_0$ is invariant under arbitrary
diffeomorphisms, this is not the case for the canonical potential $\U$.
Nevertheless, there is an important exception as under the
$GL(2,{\bf C})$--transformations
\be
q^v ={(Aq+B)/ (Cq+D)},\qquad p_v = p(Cq+D)^2/(AD-BC),
\l{4zxz}\ee
we have $\{(Aq+B)/(Cq+D),\s\}/2
=\U(\s)$, so that we can speak of the $GL(2,{\bf C})$--symmetry of the
canonical equation.
Involutivity of the Legendre transformation and the duality
$$
\S_0\longleftrightarrow \T_0,\quad ~~~{q}\longleftrightarrow p,
$$
imply another $GL(2,C)$--symmetry, with the dual versions of Eq.(\ref{10})
being
\be
\left(\partial^2_t+{\V}({\t})\right)p\sqrt q
=0=\left(\partial^2_{\tt t}+{\V}({\tt t})\right)\sqrt q,
\l{abbdual}\ee
where ${\V}({\tt t})=\{p\sqrt q/\sqrt q,{\tt t}\}/2=\{p,{\tt t}\}/2$,
with $\t=\T_0(p)$. We note that for $p=\gamma/q$ the solutions of (\ref{10})
and (\ref{abbdual}) coincide. Therefore we have the self--dual states
\be
\S_0=\gamma\ln \gamma_q q,\qquad {\T}_0=\gamma\ln \gamma_p p,
\l{selfduals}\ee
where the three constants satisfy
\be
\gamma_p\gamma_q\gamma=e.
\l{tregammae}\ee
Observe that
\be
\S_0+\T_0=pq=\gamma,\qquad \U({\tt s})=-{1/ 4\gamma^2}=\V({\tt t}).
\l{s0t0}\ee
The canonical equation (\ref{10}) and its dual (\ref{abbdual}) correspond to two
equivalent descriptions of the physical system. Remarkably, for the
self--dual
states the two descriptions overlap. Furthermore, we observe that the canonical
equation and its dual are covariant under arbitrary transformations. Actually,
under $q\longrightarrow \tilde q$, $\S_0\longrightarrow\tilde\S_0(\tilde q)$ the
transformation properties of $\T_0$ are determined by the fact that $\tilde\T_0
(\tilde p)$ is the Legendre transformation of $\tilde \S_0(\tilde q)$: $\T_0(p)
\longrightarrow \tilde\T_0(\tilde p)=\tilde p\tilde q-\tilde\S_0(\tilde q)$.
Repeating the above derivation one sees that the canonical equation and its dual
have the same form as the original ones.

The transformations in (\ref{6xxx}) and (\ref{cf1}) do not correspond to 
canonical transformations. Since $p$ and
$q$ are considered dependent, a transformation of $q$ induces a transformation
of $p$ and vice versa. Thus, in \cite{1}, as in the search for canonical
transformations leading to a system with vanishing Hamiltonian one obtains the 
HJ equation, we looked for transformations on the {\it dependent} quantities $q$
and $p=\partial_q\S_0(q)$ reducing to the free system with vanishing energy. 

The answer to this basic question led to the formulation of an equivalence
principle, suggested by the fact that $\U$, though invariant
under M\"obius transformations, changes under arbitrary diffeomorphisms. This
equivalence principle led to the quantum analogue of the HJ equation \cite{1}.
Therefore, we have the following problem: given an arbitrary reduced
action $\S_0(q)$, find the map $q\longrightarrow q^{0}=
v_0(q)$, such that the new reduced action $\S_0^{0}$, defined by
\be
\S_0^{0} (q^{0})=\S_0(q),
\l{universalstate}\ee
corresponds to the free system with vanishing energy.
Observe that the structure of the states described by $\S_0^{0}$ and $\S_0$ 
determines the ``trivializing coordinate" $q^0$ to be
\be
q\longrightarrow q^0 =\S_0^{0^{\;-1}}\circ \S_0(q),
\l{9thebasicidea}\ee

Let us set $\W\equiv V(q)-E$, where $V$ is the potential and $E$ is the energy.
We denote by $\H$ the space of all possible ${\W}$'s. If the above question has
solution then there is the following ``diffeomorphic equivalence principle'' 
\cite{1}

\vspace{.33cm}

\noindent
{\it For each pair $\W^a,\W^b\in\H$, there is a $v$--transformation such that}
\be
\W^a(q)\longrightarrow {\W^a}^v (q^v)=\W^b(q^v).
\l{equivalence}\ee
This implies that there always exists the trivializing coordinate $q^0$ for
which $\W(q)\longrightarrow \W^0(q^0)\equiv 0$. In particular,
since the inverse transformation should exist as well, it is clear that the
trivializing transformation should be a continuous, locally one--to--one map.

In \cite{1} it has been shown that this principle cannot be consistently
implemented in classical mechanics. Actually, note that the
Classical Stationary HJ Equation (CSHJE) 
\be
{1\over 2m}({\partial_{q}\S_0^{cl}(q)})^2+\W(q)=0,
\l{012}\ee 
provides a correspondence between $\W$ and $\S_0^{cl}$. In particular, $\S_0^{
cl\, v}(q^v)$ must satisfy the CSHJE $({\partial_{q^v}\S_0^{cl\, v}(q^v)})^2
/2m+\W^v (q^v)=0$. Since $\S_0^{cl\,v}(q^v)=\S_0^{cl}(q)$, by (\ref{012})
\be
\W(q)\longrightarrow \W^v(q^v)=(\partial_q q^v)^{-2}\W(q).
\l{natura2}\ee
Therefore, in classical mechanics consistency requires that $\W(q)$ belongs to
the space $\Q$ of functions transforming as quadratic differentials under
$v$--maps.

Let us now consider the case of the state $\W^0$. By (\ref{natura2}) it follows
that
\be
\W^0(q^0)\longrightarrow\W^v(q^v)=\left(\partial_{q^0}q^v\right)^{-2}\W^0(q^0)
=0.
\l{qddd}\ee
Then we have \cite{1}

\vspace{0.33cm}

\noindent
{\it In classical mechanics consistency requires that a state $\W$ transforms
as a quadratic differential under the $v$--maps. As a consequence the state
${\W^0}$ is a fixed point in ${\H}$. Equivalently, in classical mechanics
the space $\H$ cannot be reduced to a point upon factorization by the
diffeomorphisms. Hence, the equivalence principle (\ref{equivalence}) cannot be
consistently implemented in classical mechanics.}

\vspace{0.33cm}

\noindent
It is therefore clear that in order to preserve the equivalence principle we 
have
to deform the CSHJE. As we will see, this request will determine the equation
for $\S_0$. Let us discuss its general form. First of all observe that adding a
constant to $\S_0$ does not change the dynamics. Actually, 
Eqs.(\ref{y1})(\ref{y2})
are unchanged upon adding a constant to either $\S_0$ or $\T_0$. Then, the most
general differential equation that $\S_0$ should satisfy has the structure
\be
F(\S_0',\S_0'',\ldots)=0,
\l{traslazione}\ee
where $'\equiv\partial_q$. Let us write down Eq.(\ref{traslazione}) in the 
general
form
\be
{1\over 2m}\left({\partial_q \S_0(q)}\right)^2+\W(q)+Q(q)=0.
\l{aa10bbbb}\ee
The properties of $\W+Q$ under the $v$--transformations (\ref{6xxx}) are 
determined
by the transformed equation $\left({\partial_{q^v}\S_0^v(q^v)}\right)^2 /2m+\W^v
(q^v)+Q^v(q^v)=0$ that by (\ref{is1}) and (\ref{aa10bbbb}) yields
\be
\W^v(q^v)+Q^v(q^v)=\left({\partial_q q^v}\right)^{-2}\left(\W(q)+Q(q)\right),
\l{yyyxxaa10bbbb}\ee
that is 
\be
(\W+Q)\in\Q.
\l{fiof}\ee

A basic guidance in deriving the differential equation for $\S_0$ is that in 
some
limit it should reduce to the CSHJE. Therefore, in determining the structure 
of the
$Q$ term we have to take into account that in the classical limit
$Q\longrightarrow 0$.
In doing this we need some parameter which will suitably select the classical 
phase.

According to the equivalence principle, all the $\W$'s are connected by a
$v$--transformation. On the other hand, we have seen that if $\W\in\Q$,
then $\W^0$ would be a fixed point in the $\H$ space. This remark and 
Eq.(\ref{fiof}) imply
\be
\W\notin \Q,\qquad \qquad Q\notin\Q.
\l{WnotinH}\ee
Therefore, the only possible way to reach $\W^v\ne 0$ from $\W^0$, is that it
transforms with an inhomogeneous term. In particular, by
(\ref{fiof})(\ref{WnotinH}) it follows that for an arbitrary state $\W^a$
we have
\be
\W^v(q^v)=\left({\partial_{q^a} q^v}\right)^{-2}\W^a(q^a)+\Z(q^a;q^v),
\l{azzoyyyxxaa10bbbb}\ee
and
\be
Q^v(q^v)=\left({\partial_{q^a}q^v}\right)^{-2}Q^a(q^a)-\Z(q^a;q^v).
\l{azzo2yyyxxaa10bbbb}\ee
Setting $\W^a=\W^0$ in Eq.(\ref{azzoyyyxxaa10bbbb}) yields 
\be
\W^v(q^v)=\Z(q^0;q^v),
\l{ddazzoyyyxxaa10bbbb}\ee
so that, according to the equivalence principle (\ref{equivalence}), all the 
states
correspond to the inhomogeneous part in the transformation of the state $\W^0$
induced by some diffeomorphism.

Let us denote by $a,b,c,\ldots$ different $v$--transformations. Comparing
\be
\W^b(q^b)=\left({\partial_{q^b}q^a}\right)^{2}\W^a(q^a)+\Z(q^a;q^b)=
\Z(q^0;q^b),
\l{ganzate}\ee
with the same formula with $q^a$ and $q^b$ interchanged we have
\be
\Z(q^b;q^a)=-(\partial_{q^a}q^b)^{2}\Z(q^a;q^b),
\l{inparticolare}\ee
and in particular $\Z(q;q)=0$.
More generally, comparing
$$
\W^b(q^b)=\left({\partial_{q^b}q^c}\right)^{2}\W^c(q^c)+\Z(q^c;q^b)=\left({
\partial_{q^b}q^a}\right)^{2}\W^a(q^a)+\left({\partial_{q^b}q^c}\right)^{2}
\Z(q^a;q^c)+\Z(q^c;q^b),
$$
with (\ref{ganzate}) we obtain
\be
\Z(q^a;q^c)=\left({\partial_{q^c}q^b}\right)^{2}\Z(q^a;q^b)-
\left({\partial_{q^c}q^b}\right)^{2}\Z(q^c;q^b),
\l{cociclo3}\ee
which is a direct consequence of the equivalence principle.

Thus, we see that the choice of representing the state transformations by the
pullback of $\S_0$ by ${v^{-1}}^*$ is the simplest one. In particular, under
the $v$--transformations $\W$, $Q$ and $\Z(q^a;q^b)$ transform as projective
connections. We will see that Eq.(\ref{cociclo3}), that can be seen
as a cocycle condition, implies $\Z(q;\gamma(q))=0=
\Z(\gamma(q);q)$, with $\gamma$ a M\"obius transformation. As this is a crucial
step in the formulation we will analyse it in detail. Actually, it is remarkable
that besides the translations and dilatations there appears a highly non
trivial symmetry such as the inversion. 

Let us first evaluate $\Z(Aq;q)$ with $A$ a non vanishing constant. 
Since $\Z(q;q)=0$ we have
\be
\Z(Aq;q)=\sum_{n=1}^\infty a_n(q)(A-1)^n.
\l{pt1}\ee
To evaluate the $q$--dependent coefficients $a_k(q)$'s we first observe that
\be
\Z(q;A^{-1}q)=\Z(AA^{-1}q;A^{-1}q)=\sum_{n=1}^\infty a_n(A^{-1}q)(A-1)^n,
\l{pt2}\ee
which can be also evaluated by first using (\ref{inparticolare}) and then the
expansion (\ref{pt1})
\be
\Z(q;A^{-1}q)=-A^2\Z(A^{-1}q;q)=\sum_{n=1}^\infty(-1)^{n+1}a_n(q)A^{2-n}(A-1)^n.
\l{pt3}\ee
Comparing (\ref{pt2}) with (\ref{pt3}) yields $a_n(A^{-1}q)=(-1)^{n+1}A^{2-n}
a_n(q)$, that is $a_n(q)=\alpha_nq^{n-2}$ where $\alpha_{2n}=0$, 
$n\in{\bf Z}_+$;
moreover, since $\Z(q;q)=0$, we have that $\Z(Aq;q)$ is vanishing at $q=0$, 
so that $\alpha_1=0$. Therefore (\ref{pt1}) becomes
\be
\Z(Aq;q)=\sum_{n=0}^\infty \alpha_{2n+3}(A-1)^{2n+3}q^{2n+1}.
\l{pt6}\ee
To fix the $\alpha_{k}$'s we first consider $\Z(q+B,q)$ with $B$ an arbitrary
constant. We have 
\be
\Z(q+B;q)=\sum_{n=1}^\infty b_n(q)B^n,
\l{am1}\ee
which follows by $\Z(q;q)=0$.
By (\ref{inparticolare})(\ref{am1}) we have $\Z(q;q+B)=-
\Z(q+B;q)=-\sum_{n=1}^\infty b_n(q)B^n$, that compared with
$\Z(q;q+B)=\Z(q+B-B;q+B)=\sum_{n=1}^\infty b_n(q+B)(-B)^n$
yields
$b_n(q+B)=(-1)^{n+1}b_n(q)$, that is $b_{2n-1}(q)=\beta_{2n-1}$, where 
$b_{2n}=0$,
$n\in {\bf Z}_+$. Therefore (\ref{am1}) becomes
\be
\Z(q+B;q)=\sum_{n=0}^\infty \beta_{2n+1}B^{2n+1},
\l{am6}\ee
Subtracting
$\Z(-q;-q+B)=\Z(-q;q)-\Z(-q+B;q)$ from $\Z(q;q-B)=\Z(q;-q+B)-\Z(q-B;-q+B)$
gives $\Z(-q;-q+B)-\Z(q;q-B)-\Z(-q,q)-\Z(q-B;-q+B)=0$, that by (\ref{pt6}) and
(\ref{am6}) becomes
\be
2\sum_{n=0}^\infty \beta_{2n+1}B^{2n+1}-\sum_{n=0}^\infty\alpha_{2n+3}2^{2n+3}
\left(q^{2n+1}+(B-q)^{2n+1}\right)=0.
\l{am10}\ee
Since this equation must be satisfied for any $B$ and $q$, we have
\be
\beta_1=4\alpha_3, \qquad \alpha_{k>3}=0,\qquad 
\beta_{k>1}=0.
\l{qw4}\ee
Note that by (\ref{cociclo3}) $(Aq+B;Aq)=A^{-2}(Aq+B;q)
-A^{-2}(Aq;q)$. On the other hand, (\ref{qw4}) implies $(Aq+B;q)=\beta_1B$ and
$(Aq;q)=\alpha_3(A-1)^3q$, so that, as $\beta_1=4\alpha_3$, we have
\be
(Aq+B;q)=\alpha_3\left[4A^2B+(A-1)^3q\right].
\l{poss1}\ee
Now observe that 
\be
(Aq+B;q)=-A^{-2}(q;Aq+B)=-A^{-2}(A^{-1}Q-A^{-1}B;Q),
\l{osserva}\ee
where $Q=Aq+B$. By (\ref{poss1}) we have
$(A^{-1}Q-A^{-1}B;Q)=\alpha_3\left[4A^{-2}(-A^{-1}B)
+(A^{-1}-1)^3Q\right]$, so that (\ref{osserva}) becomes
$(Aq+B;q)=-\alpha_3A^2\left[4A^{-2}(-A^{-1}B)+(A^{-1}-1)^3(Aq+B)\right]$,
that compared with (\ref{poss1}) yields $\beta_1=\alpha_3=0$. Therefore
\be
\Z(Aq;q)=0=\Z(q;Aq),
\l{pt11}\ee
and
\be
\Z(q+B;q)=0=\Z(q;q+B).
\l{pt12}\ee

Eq.(\ref{cociclo3}) implies $\Z(q^a;Aq^b)=A^{-2}(\Z(q^a;q^b)-\Z(Aq^b;q^b))$
so that by (\ref{pt11}) 
\be
\Z(q^a;Aq^b)=A^{-2}\Z(q^a;q^b).
\l{porcala}\ee
By (\ref{inparticolare}) and (\ref{porcala}) we have $\Z(Aq^a;q^b)
=-(\partial_{q^b} q^a)^{2}\Z(q^b;q^a)$,
then using again (\ref{inparticolare})
\be
\Z(Aq^a;q^b)=\Z(q^a;q^b).
\l{porcalana}\ee
Likewise, by (\ref{pt12})
\be
\Z(q^a+B;q^b)=\Z(q^a;q^b)=\Z(q^a;q^b+B).
\l{porcalanazummolo}\ee
Let us set $f(q)=q^{-2}\Z(q;q^{-1})$. Eqs.(\ref{inparticolare})(\ref{porcalana})
imply $f(Aq)=-f(q^{-1})$, so that $\Z(q;q^{-1})=0=\Z(q^{-1};q)$ that together to
(\ref{cociclo3}) yields $\Z(q^a;q^{b^{-1}})=q^{b^4}\Z(q^a;q^b)$. It follows that
$$
\Z(q^{a^{-1}};q^b)=-\left(\partial_{q^b}q^{a^{-1}}\right)^2\Z(q^b;q^{a^{-1}})=
-\left(\partial_{q^b}{q^a}\right)^2\Z(q^b;q^a)=\Z(q^a;q^b).
$$
Therefore
\be
\Z(q^{a^{-1}};q^b)=\Z(q^a;q^b)=q^{b^{-4}}\Z(q^a;q^{b^{-1}}).
\l{pt22}\ee
Since dilatations, translations and inversion generate the M\"obius group,
we have by (\ref{porcala})--(\ref{pt22})
\be
\Z(\gamma(q^a);q^b)=\Z(q^a;q^b),
\l{am21}\ee
and
\be
\Z(q^a;\gamma(q^b))=(Cq^b+D)^4\Z(q^a;q^b),
\l{am22}\ee
where $\gamma(q)={Aq+B\over Cq+D}$, with $\left(\begin{array}{c}A\\C\end{array}
\begin{array}{cc}B\\D\end{array}\right)\in GL(2,{\bf C})$. We also have
\be
\Z(\gamma(q);q)=0=\Z(q;\gamma(q)).
\l{am24}\ee
The above investigation implies that $\Z(q^a;q^b)$
is proportional to $\{q^a;q^b\}$. To show this, we first note that 
(\ref{cociclo3}) is the transformation rule of the Schwarzian derivative.
Then note that the identities
${\partial_x} {h'}^{1/2}{h'}^{-1/2}=0={\partial_x} 
{h'}^{-1}{\partial_x}{h'}^{1/2}{h'}^{-1/2}h$,
imply that the second--order operator
\be
{h'}^{1/2}{\partial\over\partial x}{1\over h'}{\partial\over\partial x}
{h'}^{1/2}={\partial^2\over \partial x^2}+{1\over 2}\{h,x\},
\l{easytosee}\ee
has solutions
\be
\left({\partial^2\over\partial x^2}+{1\over 2}\{h,x\}\right){h'}^{-1/2}(Ah+B)=
0=\left({\partial^2\over\partial x^2}+{1\over 2}\{h,x\}\right){h'}^{-1/2}(Ch+D).
\l{icuhy}\ee
Therefore, the Schwarzian derivative of the
ratio of two linearly independent solutions
of $(\partial_x^2+V(x))f=0$, is twice $V(x)$. Noticing 
that for any $A$ and $B$, not simultaneously vanishing, $(\partial_x^2+V(x))
f_k(x)=0$, $k=1,2$, is equivalent to $V=-(Af_1''+Bf_2'')/(Af_1+Bf_2)$, we have
$\{\gamma(h),x\}=\{h,x\}$, which implies $\{\gamma(x),x\}
=\{x,x\}=0$. Conversely, if $\{h,x\}=0$, then solving $(\ln h'(x))''-{1\over 2}
[(\ln h'(x))']^2/2=0$, gives $h(x)=\gamma(x)$. Summarizing, we have
\be
\{f,x\}=\{h,x\},
\l{dj3eccb}\ee
if and only if $f=\gamma(h)$. Let us now solve the Schwarzian equation for $f(q^a)$
\be
\{f(q^a),q^b\}=-{4m\over\beta^2}\Z(q^a;q^b),
\l{equazione}\ee
where $\beta$ is a constant that (\ref{equazione}) fixes to have
the dimension of an action.

Eq.(\ref{am24}), which represents the core of the properties of $\Z(q^a;q^b)$
derived from (\ref{cociclo3}), is crucial to derive $f(q^a)$.
Actually $\Z(q^a;\gamma(q^a))=0$ implies $\{f(q^a),\gamma(q^a)\}=0$.
On the other hand, by (\ref{dj3eccb}) $f(q^a)=(A'q^a+B')/(C'q^a+D')$.
Therefore, we can state the central result 
\be
\Z(q^a;q^b)= -{\beta^2\over 4m} \{q^a,q^b\},
\l{equazione3}\ee
which, as we have seen, uniquely follows from the equivalence principle 
(\ref{equivalence}). Remarkably, (\ref{equazione3}) also naturally selects the
parameter leading to the classical phase. Actually, 
comparing $\W^v(q^v)=\left(\partial_q q^v\right)^{-2}\W(q)+\Z(q;q^v)$, 
with the classical version $\W^v(q^v)=\left(\partial_q q^v\right)^{-2}\W(q)$
one sees that in the classical limit
${\beta^2}\{q,q^v\}/4m\longrightarrow 0$.
Thus $\beta$ is precisely the parameter we are looking for. In particular
\be
\lim_{\beta \longrightarrow 0} Q=0,
\l{comennsorise}\ee
and $\lim_{\beta \rightarrow 0} \S_0= \S_0^{cl}$.
Eqs.(\ref{ddazzoyyyxxaa10bbbb})(\ref{equazione3}) 
imply that $\W$ itself is a Schwarzian derivative
$\W^v(q^v)=-{\beta^2\over 4m}\{q^0,q^v\}$.
We will see that the unique possible $Q$ in (\ref{aa10bbbb}) is
\be
Q={\beta^2\over 4m}\{\S_0,q\},
\l{sothat2}\ee
that by (\ref{aa10bbbb}) and the basic identity
\be
\left({\partial_q\S_0}\right)^2=\beta^2( \{e^{{2i\over\beta}\S_0}
,q\}-\{\S_0,q\})/2,
\l{expoid}\ee
is equivalent to
\be
\W=-{\beta^2\over 4m}\{e^{{2i\over\beta}\S_0},q\}.
\l{sothat}\ee
By (\ref{easytosee}) and (\ref{icuhy}) it follows that \cite{1}
\be
e^{{2i\over\beta}\S_0}={A\psi^D/\psi+B\over C\psi^D/\psi +D}, 
\l{dfgtp}\ee
where $\psi^D$ and $\psi$ are linearly independent
solutions of the stationary Schr\"odinger equation
\be
\left(-{\beta^2\over 2m}
{\partial^2\over \partial q^2}+V(q)\right)\psi=E\psi.
\l{yz1xxxx4}\ee
Thus, for the ``covariantizing parameter'' we have
\be
\beta=\hbar.
\l{Planck}\ee

To show the uniqueness of the solution (\ref{sothat}) we first set
$Q={\beta^2\over 4m}\{\S_0,q\}-g(q)$, so that by (\ref{aa10bbbb}) and
(\ref{expoid}) $\W=-{\beta^2\over 4m}\{e^{{2i\over\beta}\S_0},q\}+g(q)$.
Since $\S_0^{cl}$ does not depend on $\beta$, we have
$$
\lim_{\beta\longrightarrow 0}\left({\beta^2\over 4m}\{\S_0,q\}-g(q)\right)=
\lim_{\beta\longrightarrow 0}{\beta^2\over 4m}\{\S_0^{cl},q\}-g^{cl}(q)=-
g^{cl}(q),
$$
and by (\ref{comennsorise})
\be
g^{cl}=0.
\l{daiiije4}\ee
Note that we used $\lim_{\beta\to 0}\{\S_0,q\}=\{\lim_{\beta\to 0}\S_0,q\}=
\{\S_0^{cl},q\}$. However, it may happen that $\{\S_0^{cl},q\}$ is not
defined, e.g. in the case of the state $\W^0$ the associated
classical reduced action is a constant. In these cases one has
to consider $\lim_{\beta\to 0}\beta^2\{\S_0,q\}$. Let us then first consider
an arbitrary state $\W$ for which $\{\S_0^{cl},q\}$ can be defined.
Observe that by (\ref{azzo2yyyxxaa10bbbb}) $g(q)\in \Q$. On the other
hand, the only possible elements in $\Q$ that can be built by
means of $\S_0$ have the form
\be
g(q)={1\over 4m}\left({\partial_q \S_0}\right)^2 G(\S_0),
\l{da4}\ee
with $G(\S_0)$ an arbitrary function of $\S_0$. In other words, there is no way
to construct an element of $\Q$ by means of higher order derivatives of
$\S_0$, because these terms would break the covariance properties of $g$
(note that these arise as consistency conditions).
Furthermore, (\ref{traslazione}) implies $G(\S_0)=c$, where $c$ is a constant.
On the other hand, as by (\ref{da4}) this constant is dimensionless, it follows
by (\ref{daiiije4}) that $c=0$. Hence
\be
g=0.
\l{finale}\ee
The extension to an arbitrary state $\W$ follows from the observation
that since $g\in\Q$, we have that
it is sufficient that $g=0$ for some $\W$ in order
that (\ref{finale}) holds for all $\W\in{\H}$.

Therefore, we have seen that the equivalence principle actually uniquely
leads to the quantum analogue of the HJ equation which in
turn implies the Schr\"odinger equation.

Let us now derive the $v$--transformation $q\longrightarrow q^0$ such that
$\W\longrightarrow \W^0$. Note that,
under the transformation $q\longrightarrow
q^b=v^b(q)$, $\S_0(q)\longrightarrow\S_0^b(q^b)=\S_0(q)$, we have
\be
\{e^{{2i\over\hbar}\S_0^b(q^b)},q^b\}=\{e^{{2i\over\hbar}\S_0(q)},q^b\}=
(\partial_q
q^b)^{-2}\{e^{{2i\over\hbar}\S_0(q)},q\}-(\partial_q q^b)^{-2}\{q^b,q\},
\l{cJi2}\ee
that is $\W^b(q^b)=\left({\partial_q q^b}\right)^{-2}\W(q)+{\hbar^2\over 4m}
(\partial_q q^b)^{-2}\{q^b,q\}$. Therefore, if
\be
q^b={Ae^{{2i\over\hbar}\S_0(q)}+B\over Ce^{{2i\over\hbar}\S_0(q)}+D},
\l{dsoi2}\ee
then, according to (\ref{cJi2}), we have $\{e^{{2i\over\hbar}\S_0^b(q^b)},q^b\}
=0$. Therefore, if (\ref{dsoi2}) is satisfied, then $\W^b(q^b)=-\hbar^2\{e^{{2i
\over\hbar}\S_0^b(q^b)},q^b\}/4m$ coincides with the state $\W^0$. This implies
that $q^b$ is a M\"obius transformation of $q^0$. It follows that
the solution of the inversion problem (\ref{9thebasicidea}) is
\be
q\longrightarrow q^0={A'e^{{2i\over\hbar}\S_0(q)}+B'\over C'e^{{2i\over\hbar}
\S_0(q)}+D'}.
\l{doiqjwdw}\ee

The particular choice of the coefficients in (\ref{dfgtp}) depends on the
initial conditions of the Quantum Stationary HJ Equation (QSHJE)
(\ref{aa10bbbb}), which is a third--order differential equation. 
Since $\S_0$ should be a real function, and since it is always possible to
choose $\psi^D$ and $\psi$ to be real linearly independent solutions of
the Schr\"odinger equation, we have ($w=\psi^D/\psi$)
\be
e^{{2i\over \hbar}\S_0\{\delta\}}=e^{i\alpha} {w+i\bar\ell\over w-i\ell},
\l{KdT3}\ee
where $\delta=\{\alpha,\ell\}$, with $\alpha\in {\bf R}$ and
$\ell$ integration constants. Observe that $\S_0\ne cnst$ is equivalent
to non--degeneracy of the 
M\"obius transformation (\ref{KdT3}), i.e. ${\rm Re}\,\ell\ne 0$.
We note that Eq.(\ref{aa10bbbb}) and its solution (\ref{KdT3}) have been
investigated by Floyd \cite{Floyd}.

Let us denote by $\psi_{E}$ the Schr\"odinger wave function
associated to a given
state of energy $E$. For any fixed set of integration constants $\delta$, 
there are coefficients $A$ and $B$ such that
\be
\psi_{E}\{\delta\}={1\over \sqrt{\S_0'\{\delta\}}}\left(A e^{-{i\over 
\hbar}\S_0\{\delta\}}+Be^{{i\over \hbar}\S_0\{\delta\}}\right).
\l{popcaSCHR}\ee
Let us define
\be
\phi=\sqrt 2{e^{-i{\alpha\over 2}} (\psi^D-i\ell\psi)\over
\hbar^{1/2} |W(\ell+\bar\ell)|^{1/2}},
\l{phi1}\ee
where the Wronskian $W=\psi'\psi^D-{\psi^D}'\psi$ is a real
non vanishing constant. We have
\be
e^{{2i\over \hbar}\S_0\{\delta\}}={\bar\phi\over\phi},
\l{phi2}\ee
and since $\phi'\bar\phi-\phi{\bar\phi}'=-2iW(\ell+\bar\ell)/\hbar|W(\ell+\bar
\ell)|$, we obtain
\be
p=\epsilon|\phi|^{-2},
\l{phi3}\ee
where the value of $\epsilon=W(\ell+\bar\ell)/|W(\ell+\bar\ell)|=sgn [W(\ell+
\bar\ell)]$ fixes the direction of motion. We observe that a basic property of
$p$ is that, due to the $Q$ term, it takes real values and is not
vanishing even in the classically
forbidden regions. Exceptions concern the cases in which
there is some space region
where $\psi_E=0$, such as in the case the infinitely deep potential well.
In this case any linearly independent solution
is infinite in this region
and we have $p=0$ (this situation arises by considering a suitable limiting
procedure).

We now consider the effect of the M\"obius transformations on the wave 
function together with the properties of the trivializing map. Then we 
will discuss the case of the harmonic oscillator. In particular, we first
consider the important point of determining the transformations
$\delta\longrightarrow \delta'$ leaving the state described by $\psi_E$
invariant. To this end it is useful to write $e^{{2i\over \hbar}\S_0}$ in a
different form. First of all observe that the expression of $e^{{2i\over \hbar}
\S_0}$ in (\ref{KdT3}) can be seen as the composition of two maps. The first
one is the Cayley transformation $w\longrightarrow z=(w+i)/(w-i)\in S^1$.
Then $e^{{2i\over \hbar}\S_0}$ is obtained as the M\"obius transformation
\be
e^{{2i\over \hbar}\S_0}=\gamma_{\S_0}(z)={az+b\over cz+c},
\l{ABCDs}\ee
where, by (\ref{KdT3}), the entries $a=\bar d$, $b=\bar c$ of the matrix
$\gamma_{\S_0}$ are
\be
\gamma_{\S_0} =\left(\begin{array}{c}
e^{{i\over 2}\alpha}(1+\bar \ell)\\ e^{-{i\over 2}\alpha}(1-\ell)\end{array}
\begin{array}{cc}e^{{i\over 2}\alpha}(1-\bar\ell)\\e^{-{i\over 2}\alpha}(1+\ell)
\end{array}\right).
\l{dFF3F1}\ee

Let us now consider the moduli transformation
$\delta\longrightarrow \delta'=\{\alpha',\ell'\}$, so that
\be
e^{{2i\over \hbar}\S_0\{\delta\}}\longrightarrow e^{{2i\over \hbar}\S_0
\{\delta'\}}=\gamma'_{\S_0}(z)=e^{i\alpha'}{w+i{\bar \ell}'\over w -i \ell'},
\l{KdT3primo}\ee
where $\gamma'_{\S_0}$ is the matrix (\ref{dFF3F1}) with $\alpha$ and $\ell$
replaced by $\alpha'$ and $\ell'$ respectively. 
Note that $e^{{2i\over\hbar}\S_0\{\delta'\}}
=\tilde\gamma_{\S_0}\left(\gamma_{\S_0}(z)\right)$, where $\tilde\gamma_{\S_0}
=\gamma'_{\S_0}\gamma_{\S_0}^{-1}$. We can now determine the possible
transformations $\delta\longrightarrow \delta'$ such that $\psi_E\{\delta'\}$
describes the same state described by $\psi_E\{\delta\}$. That is,
we consider the transformations of the integration constants of
the QSHJE, corresponding to real $p$, such that $\psi_E$
remains unchanged up to some multiplicative constant $c$, that is
\be
\psi_E\{\delta\}\longrightarrow \psi_E\{\delta'\}=c\psi_E\{\delta\}.
\l{ccooss}\ee
Observe that $\psi_{E}\{\delta'\}=\left(\hbar\partial_q\gamma_{\S_0}'/2i
\right)^{-1/2}\left(A+B\gamma_{\S_0}'\right)$. Since 
$\tilde\gamma_{\S_0}=\gamma'_{\S_0}\gamma_{\S_0}^{-1}$, we have
\be
\partial_q\gamma_{\S_0}'=\partial_q \tilde\gamma_{\S_0}
\left(\gamma_{\S_0}(z)\right)={\partial\tilde\gamma_{\S_0}
\over \partial \gamma_{\S_0}}\partial_q \gamma_{\S_0}=
{\partial_q\gamma_{\S_0}\over (\tilde c\gamma_{\S_0}+\tilde d)^2},
\l{FcV6}\ee
where we used $\tilde\gamma_{\S_0}(\gamma_{\S_0})=(\tilde a\gamma_{\S_0}+
\tilde b)/(\tilde c\gamma_{\S_0}+\tilde d)$, with 
$\tilde a=\bar{\tilde d}$, $\tilde b=\bar{\tilde c}$,
denoting the elements of $\tilde\gamma_{\S_0}$ 
given by (\ref{dFF3F1}) with the $\delta$--moduli replaced by $\tilde\delta=
\{\tilde\alpha,\tilde\ell\}$. Therefore
\be
\psi_{E}(\delta')=\left(\hbar\partial_q\gamma_{\S_0}/2i\right)^{-1/2}
\left[A\tilde d+B\tilde b+(A\tilde c+B\tilde a)\gamma_{\S_0}\right],
\l{ppssid}\ee
and Eq.(\ref{ccooss}) is solved by
\be
A^2 \bar{\tilde b}+AB\tilde a=AB\bar{\tilde a}+B^2\tilde b,
\l{Kj9}\ee
which explicitly shows that there
are transformations of the $\delta$--moduli, and therefore of $\S_0$
and $p=\partial_q\S_0$, such
that the unit ray $\Psi_E$ associated to $\psi_E$ remains invariant. 
In the case in which
either $A$ or $B$ vanish, one has by (\ref{Kj9}) that
the transformations of $\S_0$ leaving $\Psi_E$ unchanged, correspond to
a phase change. However, as we will see, non--trivial transformations arise for
bounded states.

Since $e^{{2i\over \hbar}\S_0}$ takes values in $S^1$, reality of the
trivializing map and Eq.(\ref{doiqjwdw}) imply that $q^0=l(A\psi^D/\psi +B)
/(C\psi^D/\psi+D)$, where $l$ is a real constant with the dimension of a
length which can be determined together with the real coefficients $A,B,C,D$.
Let us denote by $\delta_0=\{\alpha_0,\ell_0\}$ the moduli associated to the
state $\W^0$. In this case we can choose $\psi^{D^0}=q^0$ and $\psi^0=1$. Since
the trivializing map is defined by $\S_0^0(q^0)=\S_0(q)$, by (\ref{KdT3})
we have
\be
e^{i\alpha_0}{q^0+i\bar\ell_0\over q^0-i\ell_0}=
e^{i\alpha}{w+i\bar\ell\over w-i\ell}.
\l{KdT301}\ee
Therefore, the trivializing map transforming the state $\W$ with moduli $\delta
=\{\alpha,\ell\}$, to the state $\W^0$ with moduli $\delta_0=\{\alpha_0,\ell_0\}$
is given by the real map
\be
q^0={(\ell_0e^{i\beta}+\bar\ell_0e^{-i\beta})w+i\ell_0\bar\ell e^{i\beta}-i\bar
\ell_0\ell e^{-i\beta}\over 2(\sin\beta)w+\ell e^{-i\beta}+\bar\ell e^{-i\beta}},
\l{g123s}\ee
where $\beta=(\alpha-\alpha_0)/2$. Let us consider the case in which the
functional structure of two reduced actions differs for a constant only, that is
$\S_0^a(q^a)=\S_0^b(q^a)+\hbar(\alpha_a-\alpha_b)/2$. Since $\S_0^a(q^a)=
\S_0^b(q^b)$, it follows that $p_a(q_a)=p_b(q_a)$, that is the
functional structure of $p_a$ and $p_b$ coincide. This is already clear from the
fact that $\S_0^a$ and $\S_0^b$ define the same system.
Therefore, we can set $\alpha=\alpha_0+2k \pi$ and (\ref{g123s}) becomes
\be
q^0={(\ell_0+\bar\ell_0)w+i\ell_0\bar\ell-i\bar\ell_0\ell\over \ell+\bar\ell}.
\l{qzeroo}\ee
We will call M\"obius states the states parameterized by $\ell$ associated to a
given $\W$.

Let us consider some properties of the QSHJE. First of all note that
since it contains the Schwarzian derivative of $\S_0$, it follows that
in order to be defined, the reduced action should be of class $C^2$
with $\S_0''$ differentiable. By (\ref{KdT3}) this is equivalent
to require that $\psi$ and $\psi^D$ be of class $C^1$
with $\psi'$ and ${\psi^D}'$ differentiable. Possible discontinuities
of $\psi'$ and ${\psi^D}'$, which may arise for example in the case of the
infinitely deep potential well, should be studied as limit cases.
The above properties was already expected as we required continuity and
local univalence of the trivializing map. However, the M\"obius symmetry
symmetry of the Schwarzian derivative implies that the continuity properties
should be extended to $\pm\infty$. In other words, the $v$--maps are local
homeomorphisms of the extended real line $\hat {\bf R}={\bf R}\cup\{\infty\}$
into itself. This means that as $q$ varies in $\hat{\bf R}$, $q^0$ spans the
extended real line a countable number of times. In \cite{4} we will
see that these conditions on the trivializing map, a direct consequence of the
equivalence principle, actually imply the standard results about
quantized energy spectra which follow from the conventional approach.
Since the M\"obius transformations, such as (\ref{g123s}) and (\ref{qzeroo}),
are globally univalent transformations of the extended real line into itself,
we have that the index of the trivializing map coincides with the index of $w$
\be
I[q^0]=I[w]=k.
\l{indice}\ee
Since according to Sturm theorem, between
any two consecutive zeroes of $\psi$ there exists a zero of
$\psi^D$, we have that $k$ is strictly related to the number of zeroes,
including the vanishing at infinity, of $\psi^D$ (or $\psi$)
\cite{4}.\footnote{This theorem can be seen as a duality
between $\psi^D$ and $\psi$. In this context we note that, while in the
standard approach one selects a solution of (\ref{yz1xxxx4}) as
the Schr\"odinger wave function, with the dual one being usually ignored, we
have that $\S_0$ and $p=\partial_q\S_0$ contain both $\psi^D$ and $\psi$.} 

As an example we consider the case of the harmonic oscillator.
In \cite{4} we will show that the only possible solutions
for the QSHJE which are consistent with the equivalence principle
are those corresponding to
the well--known spectrum $E_n=(n+1/2)\hbar \omega$, $n=0,1,\ldots$.
To derive the trivializing map, we note that by Wronskian arguments, it
follows that if $\psi$ is a solution of the Schr\"odinger equation, then a
linearly independent solution is given by\footnote{Note that both $\psi^D$
and $\psi$ can be written in terms of hypergeometric functions.} $\psi^D(q)
=-W\psi(q)\int^q_{q_0}dx \psi^{-2}(x)$ so that
$w=-W\int^q_{q_0}dx \psi^{-2}(x)$. We can choose $\psi$ to be the normalized
Hamiltonian eigenfuction $\psi_n=
c_n e^{-{m\omega \over 2\hbar}q^2}H_n((m\omega /\hbar)^{1/2} q)$,
with $H_n$ denoting the $n^{th}$ Hermite polynomial and 
$c_n=(m\omega/\pi\hbar)^{1/4}/\sqrt{2^nn!}$. 
For the dual solution we have $\psi^D_n=-W\psi_n\int^q_{q_0}dx
\psi_n^{-2}(x)$. Note that replacing $\psi_n^D$ and $\psi_n$
with two real linearly independent combinations, which 
is equivalent to perform a
real M\"obius transformation of $w_n=\psi^D_n/\psi_n$, 
corresponds changing $\alpha$ and $\ell$ in $\S_0$.

Since $\psi_n$ has $n$--zeroes at finite $q$ and vanishes at $\infty$, 
we have that the trivializing map for the $n^{th}$--state
of the harmonic oscillator has covering index $k=I[w]=n+1$.
As an example, note that  
$w_0=-c_0^{-2}W\int^q_{q_0} dxe^{{m\omega\over \hbar}x^2}$ is 
univalent, i.e. $w_0'\ne 0$ in ${\bf R}$, and vanishes at $q=q_0$.

Observe that by (\ref{popcaSCHR})--(\ref{phi3}) we have
$\phi=\epsilon^{1/2}e^{-{i\over \hbar}\S_0}/\sqrt p$ and
$\psi_E\{\delta\}= A\epsilon^{-1/2}\phi+ B\epsilon^{1/2}\bar\phi$
(note that $\overline{\epsilon^{-1/2}}=\epsilon^{1/2}$).   
Let us determine $A$ and $B$ for bounded states (e.g. 
the harmonic oscillator). By (\ref{phi1}) we have
\be
\psi_n={\sqrt 2\over\hbar^{1/2}
|W(\ell+\bar\ell)|^{1/2}}\left(A\epsilon^{-1/2}e^{-i\alpha/2}(\psi_n^D-i\ell \psi_n)+
B\epsilon^{1/2}e^{i\alpha/2}(\psi^D_n+i\bar\ell \psi_n)\right),
\l{daaa}\ee
so that
\be
A= i\left[{e^{i\alpha}\hbar W\over
2(\ell+\bar\ell)}\right]^{1/2},\qquad B=-\epsilon e^{-i\alpha}A=\bar A.
\l{AeB}\ee
This allows us to find the transformations of $\S_0$ leaving
invariant the unit ray. Actually, according to (\ref{Kj9})
we have that $\S_0$ and $\tilde \S_0$, with
\be
e^{{2i\over \hbar}\tilde\S_0}=
{a e^{{2i\over \hbar}\S_0}+b\over
\bar{b}e^{{2i\over \hbar}\S_0} +\bar{a}},
\l{ttrrf}\ee
where
\be
{\rm Im}\, a=-\epsilon {\rm Im}\, (e^{-i\alpha} b),
\l{aeb2}\ee
correspond to the same unit ray defined by $\psi_n$.

Let us now consider the expectation value
\be
\langle \widehat \O\rangle=\int^{\infty}_{-\infty}dq \bar \psi 
\widehat\O \psi,
\l{aspettazione}\ee
where $\psi$ is some superposition of eigenstates of the harmonic 
oscillator Hamiltonian. Since $\psi=\sum_na_n\psi_n$ and
$\widehat\O\psi=\sum_nb_n\psi_n$, we have that to consider the effect
of the trivializing map on the integrand of (\ref{aspettazione})
reduces to the problem of considering its action on $|\psi_n|^2$ for any
$n$. By (\ref{popcaSCHR}) we have that $\psi_n$ transforms as a
$-1/2$--differential under $v$--maps. In this context we note that
since there is a trivializing map for any (M\"obius) state it should be
possible to consider different coordinates for each $\psi_n$. This is just
the case as we can replace $\psi(q)=\sum_na_n\psi_n(q)$ in
(\ref{aspettazione}) with $\sum_na_n\psi_n(q_n)$ and simultaneously
changing the measure. In particular, since
$|\int_{-\infty}^{+\infty}dq\psi_n(q)|<\infty$, $\forall
n$, we have $\lim_{N\to\infty,L\to\infty}(2L)^{-N}\int^L_{-L}
(\prod_{k=0}^Ndq_k)\psi_m(q_m)\psi_n(q_n)=\delta_{nm}$, so that
\be
\langle\widehat\O\rangle=\lim_{N\to\infty,L\to\infty}(2L)^{-N}\int^L_{-L}
(\prod_{k=0}^Ndq_k)\sum_{m=0}^N\bar a_m\psi_m(q_m)\sum_{n=0}^Nb_n
\psi_n(q_n).
\l{aspettazione2}\ee
It is interesting that considering the trivializing map for an arbitrary
state leads to consider the measure $\prod_{k=0}^\infty dq_k$. 
In this context we note that very recently the transformations leading to 
the free system with 
vanishing energy have been considered by Periwal in the path--integral 
framework \cite{Periwal}.

Observe that by (\ref{KdT3}) it follows that
the two self--dual states (\ref{selfduals})
with $\gamma=\pm{i\over 2}\hbar$ 
correspond to complex M\"obius states of $\W^0$. Actually, the
difference between $\S_0^0(q^0)$ and 
$\S_0^{sd}=\pm{i\over 2}\hbar\ln\gamma_q q$
amount to a complex M\"obius transformation which has no effect on 
$\W^0=-\hbar^2\{e^{{2i\over\hbar}\S_0^0(q^0)},q^0\}/4m  =0$.

While in 
the standard approach the solution corresponding to the state $\W^0$ coincides
with the classical one, here we have a basic difference as the equivalence
principle cannot be implemented if one considers the solution $\S_0^0=cnst$.
In particular, the Schwarzian derivative of $\S_0^0$ is not defined
for $\S_0^0=cnst$. This aspect is strictly related to the existence of the
Legendre transformation of $\S_0$ for any state \cite{1}. 
 Using the identity $\{q,\S_0\}=-(\partial_q \S_0)^{-2}
\{\S_0,q\}$, one sees that the quantum correction to the CSHJE corresponds to
the conformal rescaling of the conjugate momentum \cite{1}
\be
{1\over 2m}\left({\partial\S_0\over\partial q}\right)^2
\left[1-\hbar^2\U(\S_0)
\right]+\W=0.
\l{hsxdgyij}\ee
This shows the basic role of the purely quantum mechanical state
$\W^0$ as in this case the QSHJE is solved by the overlooked 
zero--modes of the conformal factor, that is
\be
1-\hbar^2\U\left({\hbar\over 2i}\ln{q^0+i\bar\ell_0\over q^0-i\ell_0}\right)=0.
\l{vabene}\ee

Our formalism naturally fits with the brackets
\be
\left\{X,Y\right\}_{(\beta)}
\equiv{\partial\over\partial a^i}X{\left(\partial_\beta
\tau\right)^{-1}}^{ij} 
\partial_\beta{\partial \over \partial a^j}Y-{\partial\over
\partial a^i}Y{\left(\partial_\beta\tau\right)^{-1}}^{ij} \partial_\beta
{\partial\over\partial a^j}X,
\l{brackets}\ee
introduced in \cite{BOMA} in the framework of $N=2$ SYM \cite{SW}. 
In particular
\be
\{a^i,a^j\}_{(\beta)}=0=\{a^D_i,a^D_j\}_{(\beta)},\qquad
\{a^i,a^D_j \}_{(\beta)}=\delta^i_j.
\l{brackets2}\ee
We refer to \cite{BOMA} for notation in (\ref{brackets})(\ref{brackets2}).
In the one--dimensional case, setting
$$
a=\sqrt{p},\qquad a^D=q\sqrt{p},\qquad\tau={\partial_a a^D},\qquad
\partial_{\beta}={\partial_\s},
$$
the canonical equation (\ref{10}), whose canonical potential essentially
coincides with the quantum potential, has the bracket representation
\be
\{\sqrt p,q\sqrt p\}_{(\s)}=1.
\l{uno}\ee
Similarly, setting $a=\psi$, $a^D=\psi^D$, $\partial_\beta=\partial_q$, we
see that the Schr\"odinger equation (\ref{yz1xxxx4}) is equivalent to the 
bracket
\be
\{\psi,\psi^D\}_{(q)}=1,
\l{due}\ee
which matches with the formalism in \cite{2}\cite{Carroll}.
These brackets, which according to (\ref{brackets}) and (\ref{brackets2})
can be extended to higher dimensions, can be seen as the analogue of the
Poisson brackets in the case in which $p$ and $q$ are dependent. 
In this context, we also observe that the inversion formula in
\cite{3}, including its higher dimensional extension \cite{STYEY}
\be
u={i\over 4\pi b_1}\left({\cal F}-\sum_i{a^i\over 2}a_i^D\right),
\l{ext}\ee
satisfies the equation \cite{BOMA}
\be
{\cal L}_\beta u=u,
\l{hgdy}\ee
where ${\cal L}_\beta$ is a second--order modular invariant operator. In our
approach, Eq.(\ref{ext}) corresponds to the higher dimensional analogue of
the Legendre transformation of $\T_0$. The generalization of the above
$GL(2,{\bf C})$--symmetry is just the symplectic group.

We observe that in \cite{Gozzi} Gozzi showed, in the framework of the HJ
theory, that the classical ``symmetry'' associated to the Lagrangian
rescaling is broken by quantum effects with the corresponding ``anomalous''
conservation law leading to the Schr\"odinger equation.

In conclusion, we note that the equivalence principle suggests a new view of
quantum mechanics and the reexamination of its basic foundation. 
In this context we observe that
some aspects of the investigation are reminiscent of Bohm theory 
\cite{Holland}. However, there are basic differences, some of
which have been emphasized 
by Floyd  who investigated the QSHJE in \cite{Floyd},
and which  we derived from the equivalence principle \cite{1}.
As stressed in \cite{Floyd}, there is no wave guide in this approach.
Another feature is that by (\ref{phi3}) 
$p$ is a real quantity also in classically forbidden regions. 
Furthermore, unlike in Bohm theory, the classical limit 
arises in a natural way. As we will show in \cite{4}, the simple
but basic difference with respect to Bohm theory is that while
he identified the physical wave function with 
$Re^{{i\over \hbar}\S_0}$, we in general 
have that $\psi_E$ is a linear combination of 
$Re^{{i\over \hbar}\S_0}$ and $Re^{-{i\over \hbar}\S_0}$
(reality of Eq.(\ref{yz1xxxx4}) implies that
if $Re^{{i\over \hbar}\S_0}$ is a solution,
then also $Re^{-{i\over \hbar}\S_0}$ is a solution).
The  consequence is that while
in Bohm theory $\S_0$ vanishes for physical wave functions
which are real up to a possible complex constant, e.g. in the case of 
the harmonic oscillator, this is never the case in our approach. 
Actually, as we have seen, this is strictly related to the equivalence 
principle as $\S_0$ is never a constant. Then in this approach we never
have $\S_0=0$. Thus, while in Bohm theory one has to consider 
the question of recovering $\S_0^{cl}$ in the $\hbar \longrightarrow 0$
limit, this problem is completely under control and natural in this approach.

Our approach has a wide range of 
consequences, some of which will be considered in \cite{4}\cite{5}. 
Besides the existence of trajectories in the classical forbidden 
regions \cite{Floyd}\cite{4}, a remarkable aspect concerns the appearance of 
the quantized spectra from the properties of the trivializing 
map \cite{4}.

\vspace{.33cm}

\noindent
It is a pleasure to thank G. Bertoldi, G. Bonelli, E. Floyd, E. Gozzi,
and M. Tonin 
for several interesting discussions.
Work supported in part by DOE Grant No.\ DE--FG--0586ER40272 (AEF)
and by the European Commission TMR programme ERBFMRX--CT96--0045 (MM).


\begin{thebibliography}{99}

\bibitem{1} A.E. Faraggi and M. Matone, {\it Quantum Mechanics from an
Equivalence Principle}, hep-th/9705108. 
\bibitem{2} A.E. Faraggi and M. Matone, \PRL{78}{97}{163}.
\bibitem{3} M. Matone, \PLB{357}{95}{342}.
\bibitem{Carroll} R. Carroll, hep-th/9607219; 97010216; 9702138; 
\NPB{502}{97}{561}.
\bibitem{BOMA} G. Bonelli and M. Matone, \PRL{77}{96}{4712}.
\bibitem{Floyd} E.R. Floyd, 
\PRD{\bf 25}{82}{1547};
{\bf D26} (1982) 1339;
{\bf D29} (1984) 1842;
{\bf D34} (1986) 3246;
{\it Phys. Lett.} {\bf 214A} (1996) 259.
\bibitem{4} A.E. Faraggi and M. Matone, paper in preparation.
\bibitem{Periwal} V. Periwal, \PRL{80}{98}{4366}.
\bibitem{SW} N. Seiberg and E. Witten, \NPB{426}{94}{19}; \NPB{431}{94}{484}.
\bibitem{STYEY} J. Sonnenschein, S. Theisen and S. Yankielowicz, 
\PLB{367}{96}{145}.\\ T. Eguchi and S.-K. Yang, \MODA{11}{96}{131}.
\bibitem{Gozzi} E. Gozzi, \PLB{158}{85}{489}; Erratum-ibid. {\bf B386} 
(1996) 495. 
\bibitem{Holland} P.R. Holland, {\it The Quantum Theory of Motion},
Cambridge Univ. Press (1993).
\bibitem{5} G. Bertoldi, A.E. Faraggi and M. Matone, paper in preparation.
\end{thebibliography}
\end{document}